\documentstyle[12pt]{article}
\begin{document}

\def\ds{\displaystyle}
\def\beq{\begin{equation}}
\def\eeq{\end{equation}}
\def\bea{\begin{eqnarray}}
\def\eea{\end{eqnarray}}
\def\beeq{\begin{eqnarray}}
\def\eeeq{\end{eqnarray}}
\def\ve{\vert}
\def\vel{\left|}
\def\ver{\right|}
\def\nnb{\nonumber}
\def\ga{\left(}
\def\dr{\right)}
\def\aga{\left\{}
\def\adr{\right\}}
\def\lla{\left<}
\def\rra{\right>}
\def\rar{\rightarrow}
\def\nnb{\nonumber}
\def\la{\langle}
\def\ra{\rangle}
\def\ba{\begin{array}}
\def\ea{\end{array}}
\def\tr{\mbox{Tr}}
\def\ssp{{\Sigma^{*+}}}
\def\sso{{\Sigma^{*0}}}
\def\ssm{{\Sigma^{*-}}}
\def\xis0{{\Xi^{*0}}}
\def\xism{{\Xi^{*-}}}
\def\qs{\la \bar s s \ra}
\def\qu{\la \bar u u \ra}
\def\qd{\la \bar d d \ra}
\def\qq{\la \bar q q \ra}
\def\gGgG{\la g^2 G^2 \ra}
\def\q{\gamma_5 \not\!q}
\def\x{\gamma_5 \not\!x}
\def\g5{\gamma_5}
\def\sb{S_Q^{cf}}
\def\sd{S_d^{be}}
\def\su{S_u^{ad}}
\def\ss{S_s^{??}}
\def\sbp{{S}_Q^{'cf}}
\def\sdp{{S}_d^{'be}}
\def\sup{{S}_u^{'ad}}
\def\ssp{{S}_s^{'??}}
\def\sig{\sigma_{\mu \nu} \gamma_5 p^\mu q^\nu}
\def\fo{f_0(\frac{s_0}{M^2})}
\def\ffi{f_1(\frac{s_0}{M^2})}
\def\fii{f_2(\frac{s_0}{M^2})}
\def\O{{\cal O}}
\def\sl{{\Sigma^0 \Lambda}}
\def\es{\!\!\! &=& \!\!\!}
\def\ar{&+& \!\!\!}
\def\ek{&-& \!\!\!}
\def\cp{&\times& \!\!\!}
\def\se{\!\!\! &\simeq& \!\!\!}


\renewcommand{\textfraction}{0.2}    
\renewcommand{\topfraction}{0.8}

\renewcommand{\bottomfraction}{0.4}
\renewcommand{\floatpagefraction}{0.8}
\newcommand\mysection{\setcounter{equation}{0}\section}

\def\baeq{\begin{appeq}}     \def\eaeq{\end{appeq}}
\def\baeeq{\begin{appeeq}}   \def\eaeeq{\end{appeeq}}
\newenvironment{appeq}{\beq}{\eeq}
\newenvironment{appeeq}{\beeq}{\eeeq}
\def\bAPP#1#2{
 \markright{APPENDIX #1}
 \addcontentsline{toc}{section}{Appendix #1: #2}
 \medskip
 \medskip
 \begin{center}      {\bf\LARGE Appendix #1 :}{\quad\Large\bf #2}
\end{center}
 \renewcommand{\thesection}{#1.\arabic{section}}
\setcounter{equation}{0}
        \renewcommand{\thehran}{#1.\arabic{hran}}
\renewenvironment{appeq}
  {  \renewcommand{\theequation}{#1.\arabic{equation}}
     \beq
  }{\eeq}
\renewenvironment{appeeq}
  {  \renewcommand{\theequation}{#1.\arabic{equation}}
     \beeq
  }{\eeeq}
\nopagebreak \noindent}

\def\eAPP{\renewcommand{\thehran}{\thesection.\arabic{hran}}}

\renewcommand{\theequation}{\arabic{equation}}
\newcounter{hran}
\renewcommand{\thehran}{\thesection.\arabic{hran}}

\def\bmini{\setcounter{hran}{\value{equation}}
\refstepcounter{hran}\setcounter{equation}{0}
\renewcommand{\theequation}{\thehran\alph{equation}}\begin{eqnarray}}
\def\bminiG#1{\setcounter{hran}{\value{equation}}
\refstepcounter{hran}\setcounter{equation}{-1}
\renewcommand{\theequation}{\thehran\alph{equation}}
\refstepcounter{equation}\label{#1}\begin{eqnarray}}


\newskip\humongous \humongous=0pt plus 1000pt minus 1000pt
\def\caja{\mathsurround=0pt}


\title{
 {\small \begin{flushright}
\today
\end{flushright}}
       {\Large
                 {\bf
The Effects of Fourth Generation  on the double Lepton
Polarization in $B \rar K \ell^+ \ell^-$ decay
                 }
         }
      }

\author{\vspace{1cm}\\
{\small V. Bashiry$^{1}$\thanks {e-mail: bashiry@ipm.ir},\, S.M.
Zebarjad$^2$ \thanks {e-mail: S.M.Zebarjad@physics.susc.ac.ir}, F.
Falahati$^2$\thanks {e-mail: phy1g832889@shiraz.ac.ir},\,\,K.
Azizi$^3$\thanks {e-mail: kazizi@newton.physics.metu.edu.tr}\,\,,
}
\\{\small $^1$ Cyprus International University, Via Mersin 10,
Turkey }\\ {\small $^2$ Physics Department, Shiraz University,
Shiraz, 71454, Iran }\\{\small $^3$ Physics Department, Middle
East Technical University, 06531 Ankara, Turkey}}
\date{}
\begin{titlepage}
\maketitle
\thispagestyle{empty}

\begin{abstract}
This study investigates the influence of the fourth generation
quarks on the double lepton polarizations in the $B \rar K \ell^+
\ell^-$ decay. Taking $|V_{t's}V_{t'b}|\sim \{0.01-0.03\}$ with
phase about $100^\circ$, which is consistent with the $b\to
s\ell^+\ell^-$ rate and the $B_s$ mixing parameter $\Delta
m_{B_s}$, we have found out that the double lepton($\mu, \, \tau$)
polarizations are quite sensitive to the existence of fourth
generation. It can serve as an efficient tool to search for new
physics effects, precisely, to indirect search for the fourth
generation quarks($t',\, b')$.
\end{abstract}

~~~PACS numbers: 12.60.--i, 13.30.--a, 14.20.Mr
\end{titlepage}

\section{Introduction}
Although the standard model (SM) of electroweak interaction has
very successfully described all existing experimental data, it is
believed that it is a low energy manifestation of a fundamental
theory. Therefore, intensive search for physics beyond the SM is
now being performed in various areas of particle physics. One
possible extension is the SM with more than three generations.

Considering the recent experimental data, i.e., LEP II,  two
different interpretations already exist. The first one insists on
the fact the fourth generation is ruled out by these experimental
data. The second one claimes that the status of the fourth
generation is subtle \cite{Frampton}. This approach illustrates
that the experimental results make some constrains on the fourth
generation parameters, i. e., masses(fourth neutrino mass has to
be greater than the half of the Z boson mass) and
mixing\cite{Kribs:2007nz}. While an unstable neutrino with mass of
50 GeV is ruled out by LEP II bounds, the stable one may be ruled
out by dark matter direct search experiments\cite{Kribs:2007nz}.
Many authors who support the existence of fourth generation
studied those effects in various areas, for instance, Higgs and
neutrino physics, cosmology and dark
matter\cite{Polonsky}--\cite{Khlopov}.

It is known that Democratic Mass Matrix approach \cite{harari},
which is quite natural in the SM framework, favors the existence
of the fourth SM family \cite{celikel,Sultansoy:2000dm}. The main
restrictions on the new SM families come from the experimental
data on the  $\rho$ and $S$ parameters \cite{Sultansoy:2000dm}.
However, the common mass of the fourth quark ($m_{t'}$) lies
between 320 GeV and 730 GeV with respect to the experimental value
of $\rho=1.0002^{+0.0007}_{-0.0004}$ \cite{PDG}. The last value is
close to the upper limit on heavy quark masses, $m_q\leq 700$ GeV
$\approx 4m_t$, which follows from partial-wave unitarity at high
energies \cite{chanowitz}. Flavor--changing neutral current (FCNC)
$b \rar s(d) \ell^+ \ell^-$ decays provide important tests for the
gauge structure of the standard model (SM) at one--loop level.
Moreover, $b \rar s(d) \ell^+ \ell^-$ decays are also very
sensitive to the new physics beyond the SM. New physics effects
manifest themselves in rare decays in two different ways, either
through new combinations with the new Wilson coefficients or
through the new operator structure in the effective Hamiltonian,
which is absent in the SM. One of the efficient ways in
establishing new physics beyond the SM is the measurement of the
lepton polarization in the inclusive $b\rar s(d)\ell^+ \ell^-$
transition\cite{R4620} and the exclusive $B\rar
K(~K^\ast,~\rho,~\gamma)~ \ell^+ \ell^-$ decays
\cite{R4621}--\cite{bashirychin}.

In this study, we investigate the possibility of searching for new
physics in the double lepton polarization of the  $B \rar K \ell^+
\ell^-$ using the SM with fourth generation of quarks($b',\, t'$).
The fourth quark ($t'$), like $u,c,t$ quarks, contributes to the
$b \rar s(d) $ transition at the loop level. Note that the fourth
generation effects have been widely studied in baryonic and
semileptonic B decays \cite{Hou:2006jy}--\cite{Turan:2005pf}.

The main problem in the description of exclusive decays is to
evaluate the form factors, i.e., the matrix elements of the
effective Hamiltonian between initial and final hadron states. It
is obvious that in order to achieve the form factors the non
pertubative QCD approach has to be used(see for example
\cite{Aliev:2003fy}).

The sensitivity of the branching ratio and various asymmetries  to
the existence of fourth generation quarks in the  $B \rar K^\ast
\ell^+ \ell^-$ decay\cite{Bashiry:2007tt} and $\Lambda_b \rar
\Lambda \ell^+ \ell^-$
\cite{Bashiry:2007pd,Zolfagharpour:2007eh}decay are investigated
and found out that the branching ratio, lepton polarization and
forward--backward asymmetry are all very sensitive to the fourth
generation parameters ($m_{t'}$, $V_{t'b}V^*_{t's}$).
Consequently, it is natural to ask whether the double lepton
polarizations are
 sensitive to the fourth generation parameters, in the
$ B \rar K \ell^+ \ell^-$ decays. In the present work, we try to
answer this question.

The paper is organized as follows: In section 2, we try to include
the fourth generation in the the effective Hamiltonian. In section
3, the general expressions for the longitudinal, transversal and
normal polarizations of leptons are obtained. In section 4, we
examine the sensitivity of these polarizations to the fourth
generation parameters ($m_{t'}$, $V_{t'b}V^*_{t's}$ ).

\section{The matrix element for the $B \to K \ell^+ \ell^-$ decay in SM4}

The matrix element of the $B \rar K \ell^+ \ell^-$ decay at the
quark level is described by the $b \rar s \ell^+ \ell^-$
transition and the effective Hamiltonian at $O(\mu)$ scale can be
written as
 \bea\label{Hgen} {\cal H}_{eff} &=& \frac{4 G_F}{\sqrt{2}}
V_{tb}V_{ts}^\ast \sum_{i=1}^{10} {\cal C}_i(\mu) \, {\cal
O}_i(\mu)~, \eea here the full set of the operators ${\cal
O}_i(\mu)$ and the corresponding expressions for the Wilson
coefficients ${\cal C}_i(\mu)$ in the SM are given in \cite{R23}.
As has already been noted, the fourth generation is introduced in
the same way as three generations in the SM, and so new operators
do not appear clearly and the full operator set is exactly the
same as in the SM. The fourth generation changes the values of the
Wilson coefficients $C_7(\mu),~C_9(\mu)$ and $C_{10}(\mu)$, via
virtual exchange of the fourth generation up type quark
$t^\prime$. The above mentioned Wilson coefficients are modified
as: \bea\label{c4} C_7^{tot}(\mu) &=& C_7^{SM}(\mu) +
\frac{\lambda_{t'}}
{\lambda_t} C_7^{new} (\mu)~, \nnb \\
C_9^{tot}(\mu) &=& C_9^{SM}(\mu) +  \frac{\lambda_{t'}}
{\lambda_t}C_9^{new} (\mu) ~, \nnb \\
C_{10}^{tot}(\mu) &=& C_{10}^{SM}(\mu) +  \frac{\lambda_{t'}}
{\lambda_t} C_{10}^{new} (\mu)~, \eea where $\lambda_f=V_{f
b}^\ast V_{fs}$ and the last terms in these expressions describe
the contributions of the $t^\prime$ quark to the Wilson
coefficients. $\lambda_{t'}$  can be parameterized as: \bea
{\label{parameter}} \lambda_{t'}=V_{t^\prime b}^\ast V_{t^\prime
s}=r_{sb}e^{i\phi_{sb}}\eea  $C_i$'s can also be re--written in
the following form:
 \bea\lambda_t C_i \rightarrow \lambda_t
C^{SM}_i+\lambda_{t'} C^{new}_i~,\eea  The unitarity of the
$4\times4$ CKM matrix leads to
\bea\lambda_u+\lambda_c+\lambda_t+\lambda_{t'}=0.\eea\ One can
$\lambda_u=V_{ub}^\ast V_{us}$ is very small compared with the
others . Then, $\lambda_t\approx -\lambda_c-\lambda_{t'}$. And
then
 \bea \lambda_t C^{SM}_i+\lambda_{t'}
C^{new}_i=\lambda_c C^{SM}_i+\lambda_{t'} (C^{new}_i-C^{SM}_i
)\eea It is clear that for the $m_{t'}\rar m_t$ or
$\lambda_{t'}\rar 0$, $\lambda_{t'} (C^{new}_i-C^{SM}_i )$ term
vanishes, as required by the GIM mechanism.

 In deriving Eq. (\ref{c4}), we factored
out the term $V_{tb}^\ast V_{ts}$ in the effective Hamiltonian
given in Eq. (\ref{Hgen}). The explicit forms of the $C_i^{new}$
can be obtained from the corresponding expression of the Wilson
coefficients in the SM by substituting $m_t \rar m_{t^\prime}$
(see \cite{R23,R25}). If the $\hat{s}$ quark mass is neglected,
the above effective Hamiltonian leads to the following matrix
element for the $b \rar s \ell^+ \ell^-$ decay \bea\label{e1}
{\cal H}_{eff} &=& \frac{G_F\alpha_{em}}{2\sqrt{2} \pi}
 V_{tb}V_{ts}^\ast
\Bigg[ C_9^{tot} \, \bar s \gamma_\mu (1-\gamma_5) b \, \bar \ell
\gamma_\mu \ell + C_{10}^{tot} \bar s \gamma_\mu (1-\gamma_5) b \,
\bar \ell \gamma_\mu \gamma_5 \ell \nnb \\
&-& 2  C_7^{tot}\frac{m_b}{q^2} \bar s \sigma_{\mu\nu} q^\nu
(1+\gamma_5) b \, \bar \ell \gamma_\mu \ell \Bigg]~, \eea where
$q^2=(p_1+p_2)^2$ and $p_1$ and $p_2$ are the final leptons
four--momenta. The effective coefficient $C_9^{tot}$ can be
written in the following form \bea C_9^{tot} = C_9 + Y(s)~, \eea
where $s' = q^2 / m_b^2$ and the function $Y(s')$ contains the
contributions from the one loop matrix element of the four quark
operators.

In addition to the short distance contributions, $Y_{per}(s')$
receives also long distance contributions, which have their origin
in the real $c\bar c$ intermediate states, i.e., $J/\psi$,
$\psi^\prime$, $\cdots$. In the present study we neglect the long
distance contributions for the sake of simplicity.
\par Now, having  the effective Hamiltonian, describing the $b \to
s \ell^+ \ell^-$ decay at a scale $\mu\simeq m_B$, we can write
down the matrix elements for the $B \to K \ell^+ \ell^-$ decay.
The matrix element for this decay can be obtained by sandwiching
the effective Hamiltonian between $B$ and $K$ meson states; which
are parameterized in terms of form-factors which depend on the
momentum transfer squared, $q^2=(p_B-p_K)^2= (p_+-p_-)^2$. It
follows from Eq.(\ref{e1}) that in order to calculate the
amplitude of the $B \to K \ell^+ \ell^-$ decay the following
matrix elements are required;
$$\lla K\vel \bar s \gamma_\mu b \ver B \rra ,
\lla K \vel \bar s i\sigma_{\mu\nu} q^\nu b \ver B \rra , \lla K
\vel \bar s b \ver B \rra , \lla K \vel \bar s \sigma_{\mu\nu} b
\ver B \rra . $$ These matrix elements are defined as follows
\cite{R17,R20}; \bea \label{e6002} \lla K(p_{K}) \vel \bar s
\gamma_\mu b \ver B(p_B) \rra = f_+ \Bigg[ (p_B+p_K)_\mu -
\frac{m_B^2-m_K^2}{q^2} \, q_\mu \Bigg] + f_0
\,\frac{m_B^2-m_K^2}{q^2} \, q_\mu , \eea

 \bea \label{e6003} \lla K(p_{K}) \vel \bar s
\sigma_{\mu\nu}
 b \ver B(p_B) \rra = -i \, \frac{f_T}{m_B+m_K}
\Big[ (p_B+p_K)_\mu q_\nu - q_\mu (p_B+p_K)_\nu\Big]~. \eea
 Note
that the finiteness of Eq. (\ref{e1}) at $q^2=0$ is guaranteed by
assuming that $f_+(0) = f_0(0)$.
\par The matrix elements $\lla K(p_{K}) \vel \bar s i \sigma_{\mu\nu}
q^\nu b \ver B(p_B) \rra$ and $\lla K \vel \bar s b \ver B \rra$ can
be derived  from Eqs. (\ref{e6002}) and (\ref{e6003}) by multiplying
both sides of these equations by $q^\mu$ and using the equations of
motion, we get; \bea \label{e6004} \lla K(p_{K}) \vel \bar s b \ver
B(p_B) \rra & = &
f_0 \, \frac{m_B^2-m_K^2}{m_b-m_s} , \\
\label{e6005} \lla K(p_{K}) \vel \bar s i \sigma_{\mu\nu} q^\nu b
\ver B(p_B) \rra & = & \frac{f_T}{m_B+m_K} \Big[ (p_B+p_K)_\mu q^2
- q_\mu (m_B^2-m_K^2) \Big] . \eea

\par As has already been mentioned, the form-factors entering
Eqs.(\ref{e6002})-(\ref{e6005}) represent the hadronization
process. In order to calculate these form-factors information
about the nonperturbative region of QCD is required. Therefore,
for the estimation of the form-factors to be reliable, a
nonperturbative approach is needed. Among the nonperturbative
approaches, the QCD sum rule is more predictive in studying the
properties of hadrons. The form-factors appearing in the $B \to K$
transition are computed in the framework of the light cone QCD sum
rules\cite{R17,R20}. We will use the result of the work in
\cite{R20} where radiative corrections to the leading twist wave
functions and $SU(3)$ breaking effects are taken into account. As
a result, the form-factors are parameterized in the following way
\cite{R20}; \beq\label{fi}
f_i(q^2)=\frac{r_1}{1-q^2/m_1^2}+\frac{r_2}{(1-q^2/m_1^2)^2} ,
\eeq where $1=+$ or $T$, and \beq\label{f0}
f_0(q^2)=\frac{r_2}{1-q^2/m_{fit}^2} , \eeq with $m_1=5.41$GeV and
the other parameters as given in Table 1.
\begin{table}[htb]
\begin{center}
\begin{tabular}{|c||c|c|c|}
  \hline
   & $r_1$ &$r_2$ &$m_{fit}^2$ \\
  \hline \hline
 $f_+ $&$ 0.162$ &$0.173$ &$--$  \\
  $f_0$ & 0. &$ 0.33$ &37.46 \\
  $f_T$&$ 0.161 $&$0.198$& $--$  \\
  \hline
\end{tabular}
\caption{ The parameters for the form-factors of the $B \to K$
transition are given in \cite{R20}.}\label{app:table:1}
\end{center}
\end{table}

\par Using the definition of the form factors given in
Eqs.(\ref{e6002})-(\ref{e6005}), we arrive at the following matrix
element for the $B \to K \ell^+ \ell^-$ decay; \bea \label{e6006}
{\cal M}(B\rightarrow K \ell^{+}\ell^{-}) & = & \frac{G_F
\alpha_{em}}{4 \sqrt{2} \pi} V_{tb} V_{ts}^\ast \Bigg\{ \bar \ell
\gamma^\mu \ell \, \Big[
A (p_B+p_K)_\mu + B q_\mu \Big]  \\
&&\hspace{1in}+ \bar \ell \gamma^\mu \gamma_5 \ell \, \Big[ C
(p_B+p_K)_\mu  + D q_\mu \Big] \nnb . \eea

\par The functions entering Eq.(\ref{e6006}) are defined as;
\bea \label{e6007} A &=& (C_{LL}^{tot} + C_{LR}^{tot} ) f_+ +
2 (C_{BR}^{tot}+C_{SL}^{tot}) \frac{f_T}{m_B+m_{K}} , \nnb \\
B &=& (C_{LL}^{tot} + C_{LR}^{tot}) f_- -
2 (C_{BR}^{tot}+C_{SL}^{tot})\frac{f_T}{(m_B+m_{K})q^2}(m_B^2-m_K^2) , \nnb \\
C &=& (C_{LR}^{tot} - C_{LL}^{tot} ) f_+ ,\nnb \\
D &=& (C_{LR}^{tot}  - C_{LL}^{tot} ) f_- ~. \nnb\eea where \bea
C_{LL}^{tot} &=& C_9^{tot} - C_{10}^{tot} ,\,\,\,\,\, C_{LR}^{tot} = C_9^{tot} + C_{10}^{tot} ~. \nnb\\
C_{SL}^{tot}&=&-2 m_s C_7^{tot},\,\,\,\,\,C_{BR}^{tot}=-2 m_b
C_7^{tot}.\eea

Considering Eq. (\ref{e6006}), we get the following result for the
dilepton invariant mass spectrum of decay rate: \bea \label{e6008}
\frac{d\Gamma}{d\hat{s}}(B \rar K \ell^+ \ell^-) = \frac{G^2
\alpha^2 m_B}{2^{14} \pi^5} \vel V_{tb}V_{ts}^\ast \ver^2
\lambda^{1/2}(1,\hat{r}_K,\hat{s}) v \Delta(\hat{s})~, \eea where
$\lambda(1,\hat{r}_K,\hat{s})=1+\hat{r}_K^2+\hat{s}^2-2\hat{r}_K-2\hat{s}-
2\hat{r}_K\hat{s}$, $\hat{s}=q^2/m_B^2$, $\hat{r}_K=m_K^2/m_B^2$,
$\hat{m}_\ell=m_\ell/m_B$, $v=\sqrt{1-4\hat{m}_\ell^2/\hat{s}}$ is
the final lepton velocity, and $\Delta(\hat{s})$ is \bea
\Delta(\hat{s})&=&\nonumber\frac{4m_B^2}{3}Re[24
m_{B}^2\hat{m}_{l}^2(1-\hat{r}_{K})D^{\star}C+\lambda
m_{B}^2(3-v^2)|A|^2+12m_{B}^2\hat{m}_{l}^2\hat{s}|D|^2\\
&+&
m_{B}^2|C|^2\{2\lambda-(1-v^2)(2\lambda-3(1-\hat{r}_{K})^2)\}]\eea
\section{Double-Lepton Polarization}
In this section, we will calculate the double--polarization
asymmetries, i.e., when polarizations of both leptons have to be
simultaneously measured. One can introduce a spin projection
operator as follows: \bea
\Lambda_1 \es \frac{1}{2} (1+\gamma_5\!\!\not\!{s}_i^-)~,\nnb \\
\Lambda_2 \es \frac{1}{2} (1+\gamma_5\!\!\not\!{s}_i^+)~, \nnb \eea
for lepton $\ell^-$ and antilepton $\ell^+$, where $i=L,N,T$
correspond to the longitudinal, normal and transversal
polarizations, respectively. Firstly, we must  define the orthogonal
vectors $s$ in the rest frame of leptons(where its vector is the
polarization vector of the lepton): \bea \label{e6010} s^{-\mu}_L
\es \ga 0,\vec{e}_L^{\,-}\dr =
\ga 0,\frac{\vec{p}_-}{\vel\vec{p}_- \ver}\dr~, \nnb \\
s^{-\mu}_N \es \ga 0,\vec{e}_N^{\,-}\dr = \ga
0,\frac{\vec{p}_K\times
\vec{p}_-}{\vel \vec{p}_K\times \vec{p}_- \ver}\dr~, \nnb \\
s^{-\mu}_T \es \ga 0,\vec{e}_T^{\,-}\dr = \ga 0,\vec{e}_N^{\,-}
\times \vec{e}_L^{\,-} \dr~, \nnb \\
s^{+\mu}_L \es \ga 0,\vec{e}_L^{\,+}\dr =
\ga 0,\frac{\vec{p}_+}{\vel\vec{p}_+ \ver}\dr~, \nnb \\
s^{+\mu}_N \es \ga 0,\vec{e}_N^{\,+}\dr = \ga
0,\frac{\vec{p}_K\times
\vec{p}_+}{\vel \vec{p}_K\times \vec{p}_+ \ver}\dr~, \nnb \\
s^{+\mu}_T \es \ga 0,\vec{e}_T^{\,+}\dr = \ga 0,\vec{e}_N^{\,+}
\times \vec{e}_L^{\,+}\dr~, \eea here $\vec{p}_\mp$ and
$\vec{p}_K$ are the three--momenta of the leptons $\ell^\mp$ and K
meson in the center of mass frame (CM) of $\ell^- \,\ell^+$
system, respectively.

The longitudinal unit vectors are boosted to the CM frame of $\ell^-
\ell^+$ by Lorenz transformation: \bea \label{e6011} \ga s^{-\mu}_L
\dr_{CM} \es \ga \frac{\vel\vec{p}_- \ver}{m_\ell}~,
\frac{E \vec{p}_-}{m_\ell \vel\vec{p}_- \ver}\dr~,\nnb \\
\ga s^{+\mu}_L \dr_{CM} \es \ga \frac{\vel\vec{p}_-
\ver}{m_\ell}~, -\frac{E \vec{p}_-}{m_\ell \vel\vec{p}_-
\ver}\dr~, \eea while the other two vectors remain unchanged.

We can now define the double--lepton polarization asymmetries as
in \cite{Fukae}: \bea \label{e6012} P_{ij}(\hat{s}) =
\frac{\ds{\Bigg(
\frac{d\Gamma}{d\hat{s}}(\vec{s}_i^-,\vec{s}_j^+)}-
\ds{\frac{d\Gamma}{d\hat{s}}(-\vec{s}_i^-,\vec{s}_j^+) \Bigg)} -
\ds{\Bigg( \frac{d\Gamma}{d\hat{s}}(\vec{s}_i^-,-\vec{s}_j^+)} -
\ds{\frac{d\Gamma}{d\hat{s}}(-\vec{s}_i^-,-\vec{s}_j^+)\Bigg)}}
{\ds{\Bigg( \frac{d\Gamma}{d\hat{s}}(\vec{s}_i^-,\vec{s}_j^+)} +
\ds{\frac{d\Gamma}{d\hat{s}}(-\vec{s}_i^-,\vec{s}_j^+) \Bigg)} +
\ds{\Bigg( \frac{d\Gamma}{d\hat{s}}(\vec{s}_i^-,-\vec{s}_j^+)} +
\ds{\frac{d\Gamma}{d\hat{s}}(-\vec{s}_i^-,-\vec{s}_j^+)\Bigg)}}~,
\eea where $i,j=L,~N,~T$, and the first subindex $i$ corresponds
to lepton while the second subindex $j$ corresponds to antilepton,
respectively.

Now, regarding the  aforementioned definitions, after doing the
straight forward calculations we obtain the following results for
$P_{ij}(\hat{s})$:
\begin{eqnarray}
P_{LL}&=&\frac{-4m_B^2}{3\Delta}Re[-24m_B^2\hat{m}_{l}^2(1-\hat{r}_{K})C^{\star}D+\lambda
m_B^2(1+v^2)|A|^2\\&-& 12m_B^2 \hat{m}_{l}^2\hat{s}|D|^2
+m_B^2|C|^2(2\lambda-(1-v^2)(2\lambda+3(1-\hat{r}_{K})^2))],\\
P_{LN}&=&\frac{-4\pi m_{B}^{3} \sqrt{\lambda
\hat{s}}}{\hat{s}\Delta}
Im[-m_{B}\hat{m}_{l}\hat{s}A^{\star}D -m_{B}\hat{m}_{l}(1-\hat{r}_{K})A^{\star}C],\\
P_{NL}&=&-P_{LN},\\
P_{LT}&=&\frac{4\pi
m_{B}^3\sqrt{\lambda\hat{s}}}{\hat{s}\Delta}Re[m_{B}\hat{m}_{l}v(1-\hat{r}_{K})|C|^2+m_{B}\hat{m}_{l}v\hat{s}C^{\star}D],\\
P_{TL}&=&P_{LT},\\
P_{NT}&=&-\frac{8m_B^2v}{3\Delta}Im[2\lambda m_{B}^2A^{\star}C],\\
P_{TN}&=&-P_{NT},\\
P_{TT}&=&\nonumber\frac{4m_B^2}{3\Delta}Re[-24
m_{B}^2\hat{m}_{l}^2(1-\hat{r}_{K})C^{\star}D-\lambda
m_{B}^2(1+v^2)|A|^2-12m_{B}^2\hat{m}_{l}^2\hat{s}|D|^2\\&+&m_{B}^2|C|^2\{2\lambda-(1-v^2)(2\lambda+3(1-\hat{r}_{K})^2)\}],\\
P_{NN}&=&\nonumber\frac{4m_B^2}{3\Delta}Re[24
m_{B}^2\hat{m}_{l}^2(1-\hat{r}_{K})C^{\star}D-\lambda
m_{B}^2(3-v^2)|A|^2+12m_{B}^2\hat{m}_{l}^2\hat{s}|D|^2\\&+&m_{B}^2|C|^2\{2\lambda-(1-v^2)(2\lambda-3(1-\hat{r}_{K})^2)\}]\eea

 \section{Numerical analysis}

In this section, we will analyze the dependence of the
double--lepton polarizations on the fourth quark mass($m_{t'}$)
and the product of quark mixing matrix elements ($V_{t^\prime
b}^\ast V_{t^\prime s}=r_{sb}e^{i\phi_{sb}}$). The challenging
input parameters in the calculations are the form factors, which
are related to the non--pertubative part of QCD.
 We will use the result of the study in
\cite{R20} where radiative corrections to the leading twist wave
functions and $SU(3)$ breaking effects are taken into account.

We use the next--to--leading order logarithmic approximation for
the resulting values of the Wilson coefficients $C_9^{eff},~C_7$
and $C_{10}$ in the SM \cite{R46215,R46216} at the
re--normalization point $\mu=m_b$. It should be noted that in
addition to the short distance contribution, $C_9^{eff}$ receives
 long distance contributions from the real $\bar c c$ resonant
states of the $J/\psi$ family. In the present research, we do not
take  the long distance effects into account.

The input parameters used in this
analysis are as follows:\\
$\vel V_{tb} V_{ts}^\ast\ver = 0.0385$, $(C_9^{eff})^{sh}= 4.344$,
$C_{10}=-4.669$, $\Gamma_B = 4.22\times 10^{-13}~GeV$.
 In order to perform quantitative analysis of the double--lepton polarizations, the values of
the new parameters($m_{t'},\,r_{sb},\,\phi_{sb}$) are needed.
Using the experimental values of $B\rar X_s \gamma$ and $B\rar X_s
\ell^+ \ell^-$, the restriction on $r_{sb}\sim\{0.01-0.03\}$ has
been obtained \cite{Arhrib:2002md,Zolfagharpour:2007eh} for
$\phi_{sb}\sim\{0-2\pi\}$ and $m_{t'}\sim\{200,600\}~$(GeV)(see
table 2). Considering the $B_s$ mixing, which is in terms of the
$\Delta m_{B_s}$, $\phi_{sb}$ is sharply restricted
($\phi_{sb}\sim\pi/2$) \cite{Hou:2006jy}.
\begin{table}
\renewcommand{\arraystretch}{1.5}
\addtolength{\arraycolsep}{3pt}
$$
\begin{array}{|c|c|c|c |c|}
\hline  r_{sb} & 0.005 & 0.01 &0.02 & 0.03 \\
\hline
m_{t'}(GeV) & 739 &529 & 385 & 331\\
\hline
\end{array}
$$
\caption{The  experimental limit of $ m_{t'} $ for
$\phi_{sb}=\pi/3$\cite{Zolfagharpour:2007eh}}
\renewcommand{\arraystretch}{1}
\addtolength{\arraycolsep}{-3pt}
\end{table}

\begin{table}
\renewcommand{\arraystretch}{1.5}
\addtolength{\arraycolsep}{3pt}
$$
\begin{array}{|c|c|c|c |c|}
\hline  r_{sb} & 0.005 & 0.01 &0.02 & 0.03 \\
\hline
m_{t'}(GeV) &511 &373 & 289 & 253\\
\hline
\end{array}
$$
\caption{The experimental limit of $ m_{t'} $ for
$\phi_{sb}=\pi/2$\cite{Zolfagharpour:2007eh}}
\renewcommand{\arraystretch}{1}
\addtolength{\arraycolsep}{-3pt}
\end{table}

\begin{table}
\renewcommand{\arraystretch}{1.5}
\addtolength{\arraycolsep}{3pt}
$$
\begin{array}{|c|c|c|c |c|}
\hline  r_{sb} & 0.005 & 0.01 &0.02 & 0.03 \\
\hline
m_{t'}(GeV) &361 &283 & 235 & 217\\
\hline
\end{array}
$$
\caption{The experimental limit of $ m_{t'} $ for
$\phi_{sb}=2\pi/3$\cite{Zolfagharpour:2007eh}}
\renewcommand{\arraystretch}{1}
\addtolength{\arraycolsep}{-3pt}
\end{table}

Before performing numerical analyses, we would like to add a few
words about lepton polarizations. From explicit expressions of the
lepton polarizations one can easily see that they depend on both
$\hat{s}$ and the new parameters($m_{t'},\,r_{sb}$). Therefore, it
may experimentally be difficult to study these dependencies at the
same time. For this reason, we eliminate the $q^2$ dependence by
performing integration over $\hat{s}$ in the allowed region, i.e.,
we consider the averaged double--lepton polarization asymmetries.
The average gained, here, over $\hat{s}$ is defined as: \bea \la
P_{ij} \ra = \frac{\ds \int_{4
\hat{m}_\ell^2}^{(1-\sqrt{\hat{r}_K})^2} P_{ij} \frac{d{\cal B}}{d
\hat{s}} d \hat{s}} {\ds \int_{4
\hat{m}_\ell^2}^{(1-\sqrt{\hat{r}_K})^2} \frac{d{\cal B}}{d
\hat{s}} d \hat{s}}~.\nnb \eea Our quantitative analyses indicate
that some of the $\la P_{ij}\ra$ are less sensitive to the fourth
generation parameters; i.e, the maximum deviation from the SM3 are
$\sim 5\%$. We do not present those dependencies on fourth
generation parameters with relevant figures. We present our
analysis for strongly dependent functions in a series of figures
where the black "DOTS" in figures show the experimental limit on
$m_{t'}$, considering the $1\sigma$ level deviation from the
measured branching ratio of $B\rightarrow X_s \ell^-\ell^+$(see
Table 2,3,4). From these figures, we deduce the following results:

\begin{itemize}
\item{ Taking the fourth generation into account , the value of
$\la P_{LL}\ra$ shows weak dependency for $\mu$ channel and alters
$\tau$ channel at most about $20\%$ compared to the SM3 the
prediction, while it increases for ,
$\phi_{sb}=90^\circ,\,120^\circ$. However, it both increases and
decreases for the $\phi_{sb}=60^\circ$ for both channels.
     }

\item{ In the SM3 the non--zero values of $\la P_{LN}\ra$ and $\la
P_{NT}\ra$, as well as $\la P_{NL}\ra$ and $\la P_{TN}\ra$, have
their origin in the higher order QCD corrections to the
$C_9^{eff}$. Since these functions are proportional to the lepton
mass and imaginary part of the $C_9^{eff}$, results of both are
negligible. However, they seem to exceed the SM value sizeably.
This is because of the new weak phase and new contribution to the
Wilson coefficients coming out of the fourth generation.
Furthermore, for fixed values of $r_{sb}$, their magnitude
decreases by increasing the $\phi_{sb}$ in the the experimentally
allowed region. }

\item{ Regarding the fourth generation, the value of $\la
P_{NN}\ra$ changes 3--4 times for $\mu$ channel and at most about
$25\%$ for $\tau$ channel compared with the SM3 prediction, while
it increases for , $\phi_{sb}=90^\circ,\,120^\circ$. But, it is
both increases and decreases for the $\phi_{sb}=60^\circ$ in both
channels.
     }
\item{ The situation for $\la P_{TT}\ra$ is similar to the  $\la
P_{NN}\ra$, if the $\mu$ channel is considered . But  $\tau$
channel depicts weak dependence on the  fourth generation
parameters(at most $\sim 5\%$ deviation from the SM3 predictions).
     }
 \item{ The value of $\la
P_{LT}\ra$ changes about two times for $\mu$ channel and at most
about $5\%$ for $\tau$ channel compared with the SM3 prediction,
while it increases for , $\phi_{sb}=90^\circ,\,120^\circ$. But, it
 both increases and decreases for the $\phi_{sb}=60^\circ$ in
both channels.
     }
\end{itemize}

Finally, let us briefly discuss whether it is possible to measure
 the lepton polarization asymmetries in experiments or not.
A required number of the events (i.e., the number of $B \bar{B}$
pair) in terms of the branching ratio ${\cal B}$ at $n \sigma$
level, $\la P_{ij} \ra$ and the efficiencies of the leptons $s_1$
and $s_2$  are given by the expression \bea N = \frac{n^2}{{\cal B}
s_1 s_2 \la P_{ij} \ra^2}~,\nnb \eea  Typical values of the
efficiencies of the $\tau$--leptons range from $50\%$ to $90\%$ for
their various decay modes\cite{R6016}. It should be noted, here,
that the error in $\tau$--lepton polarization is estimated to be
about $(10 \div 15)\%$ \cite{R6017}. So, the error in measurement of
the $\tau$--lepton asymmetries is approximately $(20 \div 30)\%$,
and the error in obtaining the number of events is about $50\%$.

Looking at the expression of  $N$, it can be understood that in
order to detect the lepton polarization asymmetries in the $\mu$ and
$\tau$ channels at $3\sigma$ level, the minimum number of required
events are (for the efficiency of $\tau$--lepton we take $0.5$):

\begin{itemize}
\item for $B \rar K \mu^+ \mu^-$ decay \bea N = \left\{
\begin{array}{ll}
3.5 \times 10^{7}  & (\mbox{\rm for} \lla P_{LL} \rra,\lla P_{LT} \rra)~,\\
5.0 \times 10^{8}  & (\mbox{\rm for} \lla P_{TL} \rra)~,\\
2.0 \times 10^{11} & (\mbox{\rm for} \lla P_{LN}
\rra)~,\end{array} \right. \nnb \eea

\item for $B \rar K \tau^+ \tau^-$ decay \bea N = \left\{
\begin{array}{ll} (1.0 \pm 0.5) \times 10^{9}  & (\mbox{\rm for}
\lla P_{LL} \rra,\lla P_{LT}
\rra,\lla P_{TL} \rra,\lla P_{NN} \rra)~,\\
(5.0 \pm 2.5) \times 10^{8}  & (\mbox{\rm for} \lla P_{TT} \rra)~,\\
(4.0 \pm 2.0) \times 10^{10} & (\mbox{\rm for} \lla P_{LN} \rra,
\lla P_{NL} \rra)~,\\
(3.0 \pm 1.5) \times 10^{11} & (\mbox{\rm for} \lla P_{NT} \rra,
\lla P_{TN} \rra)~.\end{array} \right. \nnb \eea
\end{itemize}

On the other hand, the number of $B \bar{B}$ pairs, that are
produced at  LHC \, are about $\sim 10^{12}$. As a result of the
comparison of these numbers and $N$, we conclude that except $\lla
P_{LN} \rra$ in the $B \rar K \mu^+ \mu^-$ decay and $\lla P_{NT}
\rra$, $\lla P_{TN} \rra$  in the $B \rar K \tau^+ \tau^-$ decay,
all double lepton polarizations can  be detectable at LHC. The
numbers for the $B \rar K \mu^+ \mu^-$ decay presented above
demonstrate that $\lla P_{LL} \rra$ and $\lla P_{LT} \rra$ for the
$B \rar K \mu^+ \mu^-$ decay might be accessible to $B$ factories
after several years of running.

To sum up, in this study we present the most general analyses of
the double--lepton polarization asymmetries in the $B \rar K
\ell^+ \ell^-$ decay using the SM with the fourth generation of
quarks. In our analyses, we have used the experimental results of
the branching ratio for the $B \rar X_s \mu^+ \mu^-$ decay and
$B_s$ mixing to control the fourth generation parameters. We have
found out that some of the double--lepton polarization functions
which are already accessible to LHC depict the strong dependency
on the fourth generation quark mass and product of quark mixing.
The study of such strong dependent double--lepton polarization
asymmetries can serve as a good test for the predictions of the SM
and for the indirect search for the fourth generation up type
quarks $t'$.

\section{Acknowledgment}
The authors would like to thank T. M. Aliev for his useful
discussions.

\newpage

\section*{Figure captions}

{\bf Fig. (1)} The dependence of the $\lla P_{LL}\rra$ for the $B
\rar K \tau^+ \tau^-$ decay on the fourth generation quark mass
$m_{t'}$ for three different values of
 $\phi_{sb}=\{60^\circ,~ 90^\circ, ~120^\circ\}$ and $r_{sb}=\{0.01,~0.02,~0.03\}$ .\\ \\
{\bf Fig. (2)} The dependence of the $\lla P_{LN}\rra$  on the
fourth generation quark mass $m_{t'}$ for three different values
of $\phi_{sb}\sim\{60^\circ,~ 90^\circ, ~120^\circ\}$ and $r_{sb}=\{0.01,~0.02,~0.03\}$  for $\mu$ lepton.\\ \\
{\bf Fig. (3)}The same as in Fig. (2), but for the $\tau$ lepton.\\ \\
{\bf Fig. (4)} The dependence of the $\lla P_{NT}\rra$  on the
fourth generation quark mass $m_{t'}$ for three different values
of $\phi_{sb}\sim\{60^\circ,~ 90^\circ, ~120^\circ\}$ and $r_{sb}=\{0.01,~0.02,~0.03\}$ for $\mu$ lepton.\\ \\
{\bf Fig. (5)}The same as in Fig. (4), but for the $\tau$ lepton.\\ \\
{\bf Fig. (6)} The dependence of the $\lla P_{LT}\rra$  on the
fourth generation quark mass $m_{t'}$ for three different values
of $\phi_{sb}\sim\{60^\circ,~ 90^\circ, ~120^\circ\}$ and $r_{sb}=\{0.01,~0.02,~0.03\}$  for $\mu$ lepton.\\ \\
{\bf Fig. (7)} The dependence of the $\lla P_{NN}\rra$  on the
fourth generation quark mass $m_{t'}$ for three different values
of $\phi_{sb}\sim\{60^\circ,~ 90^\circ, ~120^\circ\}$ and $r_{sb}=\{0.01,~0.02,~0.03\}$  for $\mu$ lepton.\\ \\
  {\bf Fig. (8)}The same as in Fig. (7), but for the $\tau$ lepton.\\ \\
{\bf Fig. (9)} The dependence of the $\lla P_{TT}\rra$  on the
fourth generation quark mass $m_{t'}$ for three different values
of  $\phi_{sb}\sim\{60^\circ,~ 90^\circ, ~120^\circ\}$ and $r_{sb}=\{0.01,~0.02,~0.03\}$  for $\mu$ lepton.\\ \\
\newpage
\begin{figure}
\vskip 1.5 cm
    \includegraphics{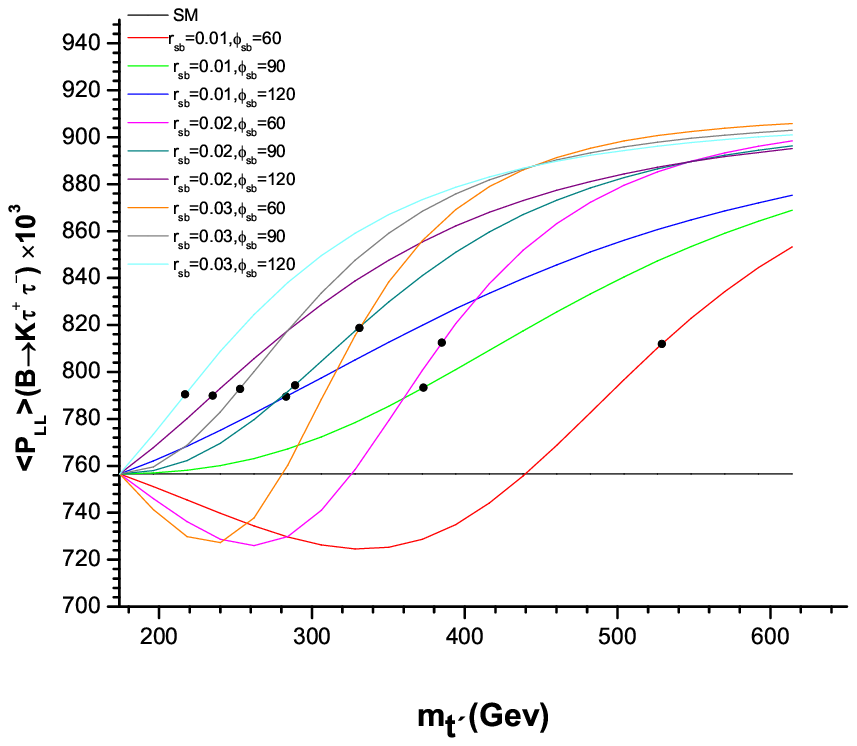}
\vskip 7.8cm \caption{}
\end{figure}

\begin{figure}
\vskip 2.5cm
    \includegraphics{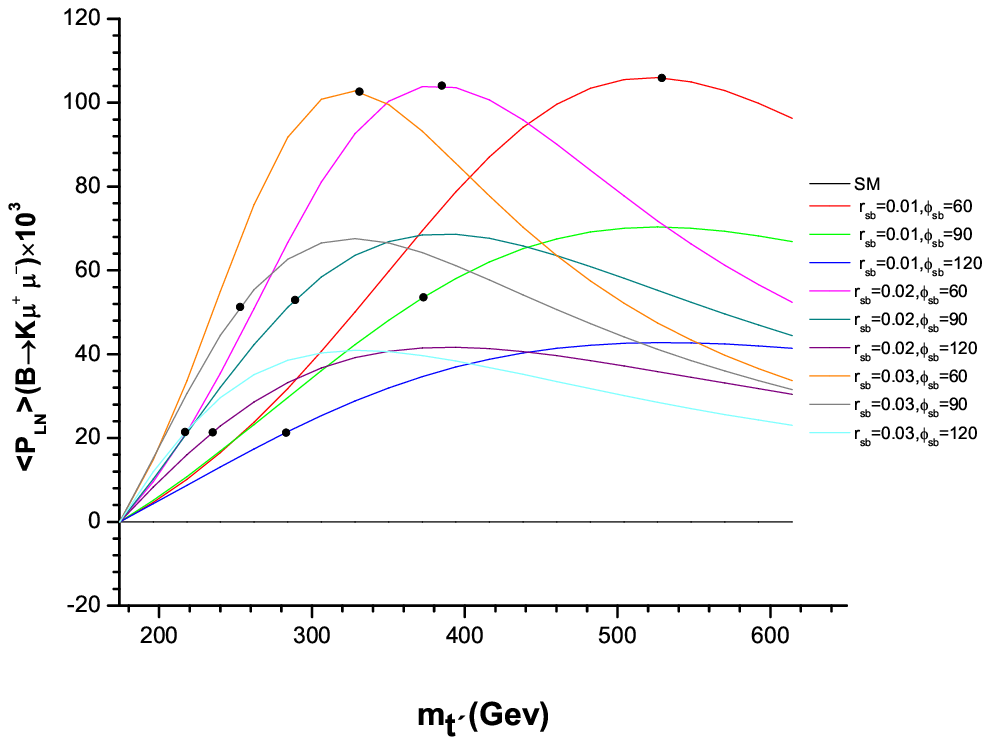}
\vskip 7.8cm \caption{}
\end{figure}

\begin{figure}
\vskip 2.5 cm
    \includegraphics{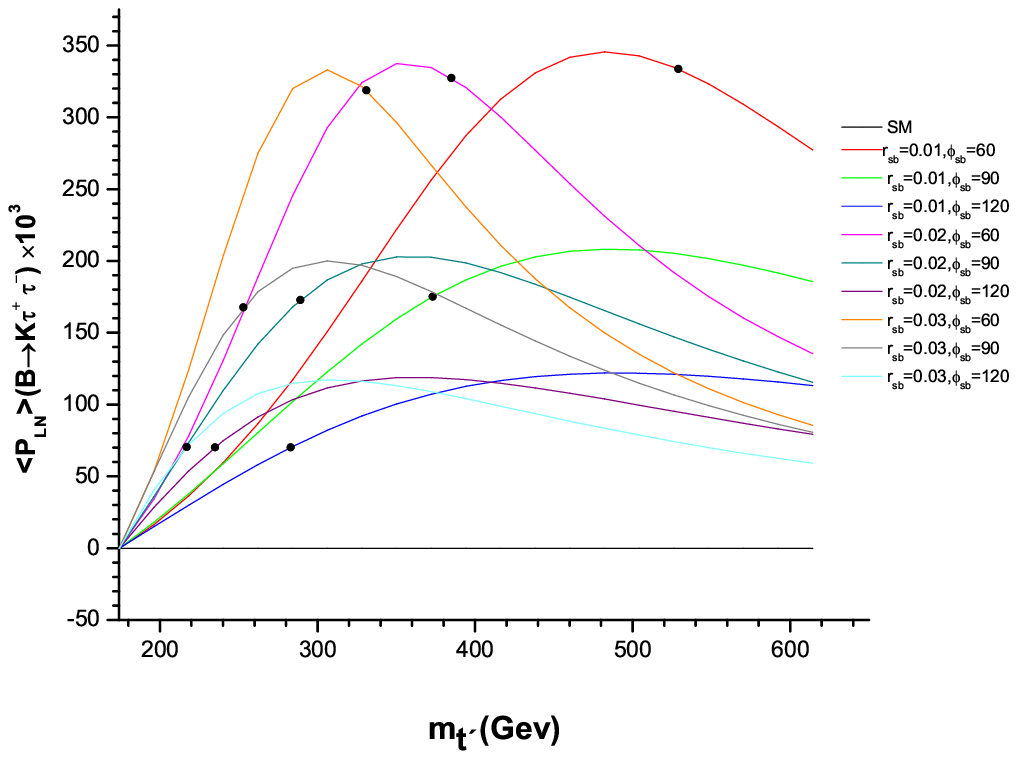}
\vskip 7.8cm \caption{}
\end{figure}

\begin{figure}
\vskip 1.5 cm
    \includegraphics{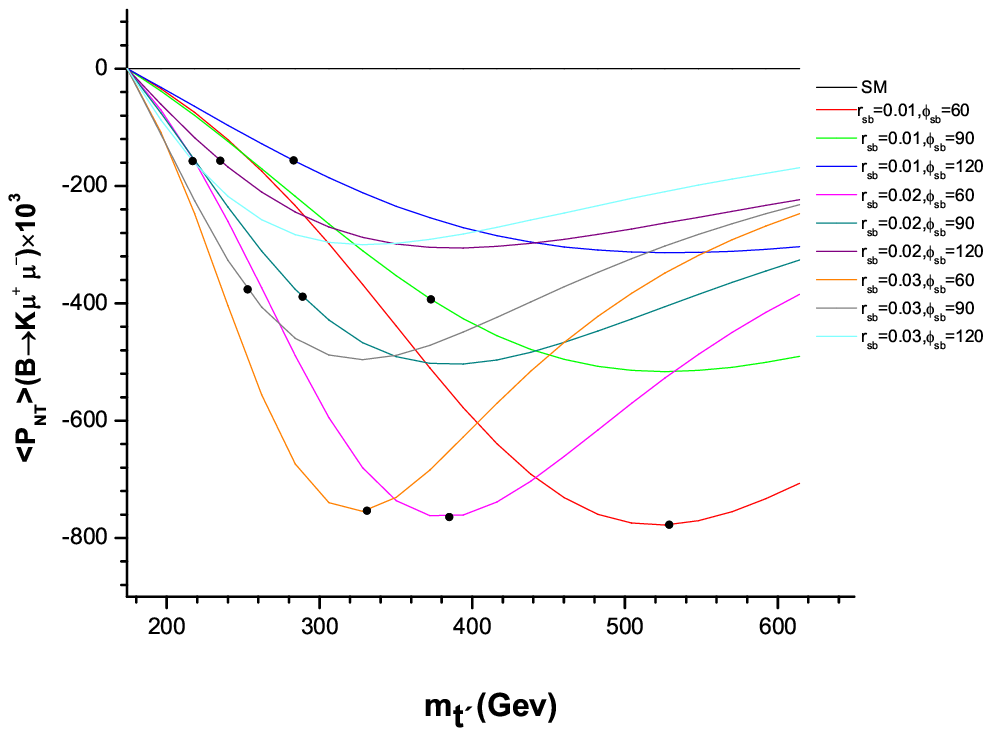}
\vskip 7.8 cm \caption{}
\end{figure}

\begin{figure}
\vskip 2.5 cm
    \includegraphics{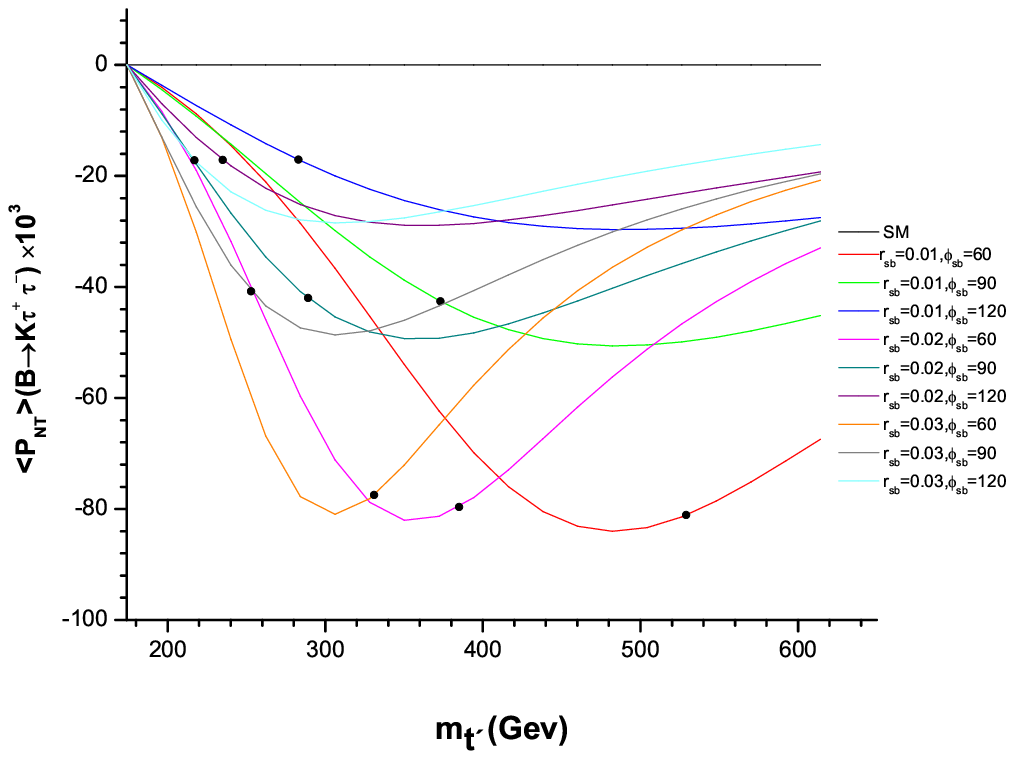}
\vskip 7.8cm \caption{}
\end{figure}

\begin{figure}
\vskip 1.5 cm
    \includegraphics{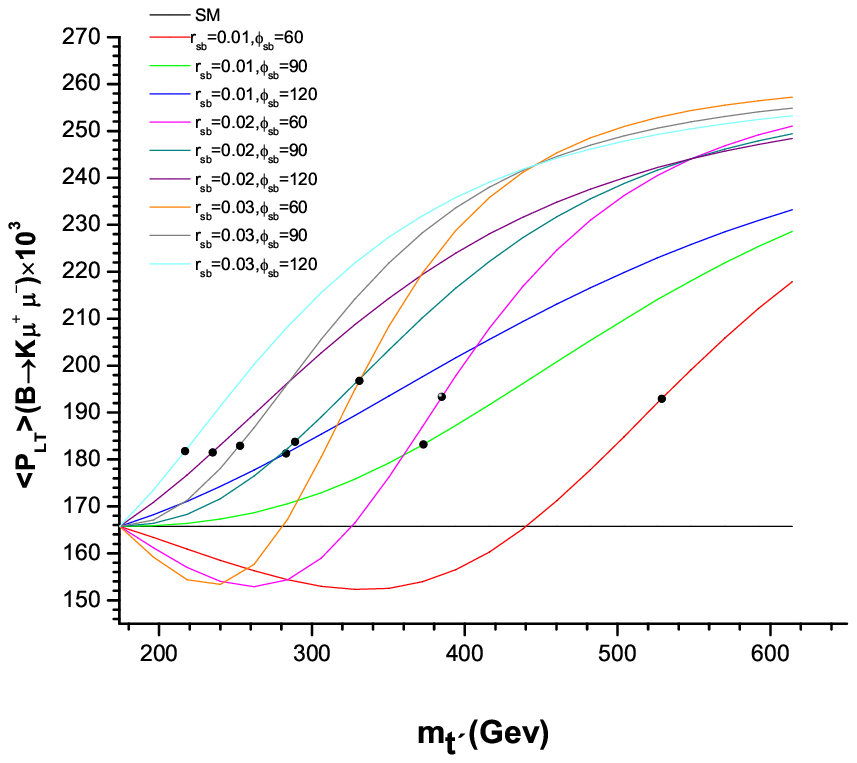}
\vskip 7.8cm \caption{}
\end{figure}

\begin{figure}
\vskip 2.5 cm
    \includegraphics{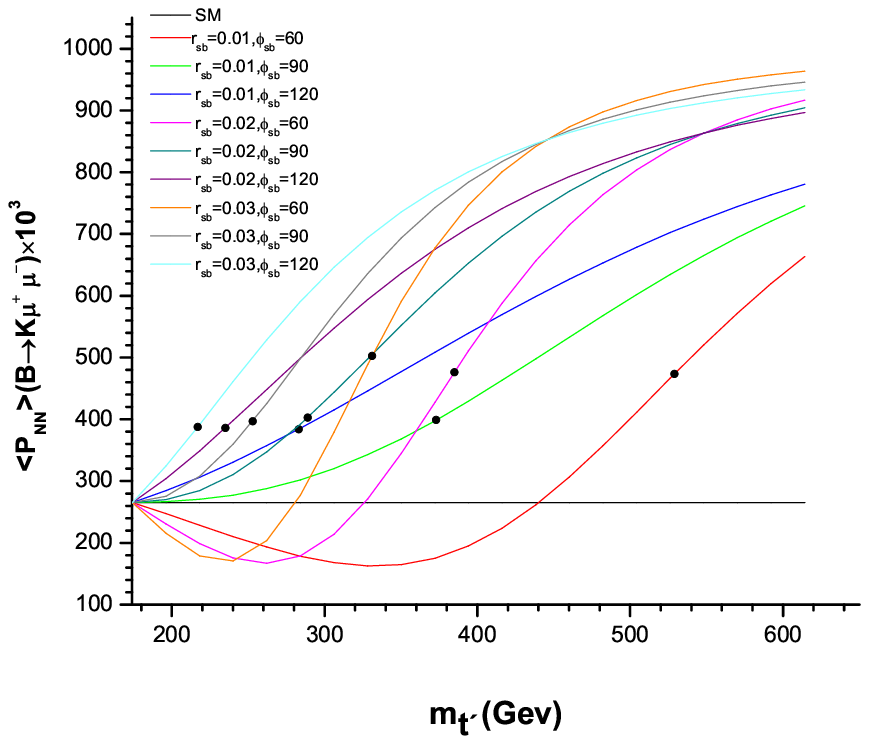}
\vskip 7.8 cm \caption{}
\end{figure}

\begin{figure}
\vskip 1.5 cm
    \includegraphics{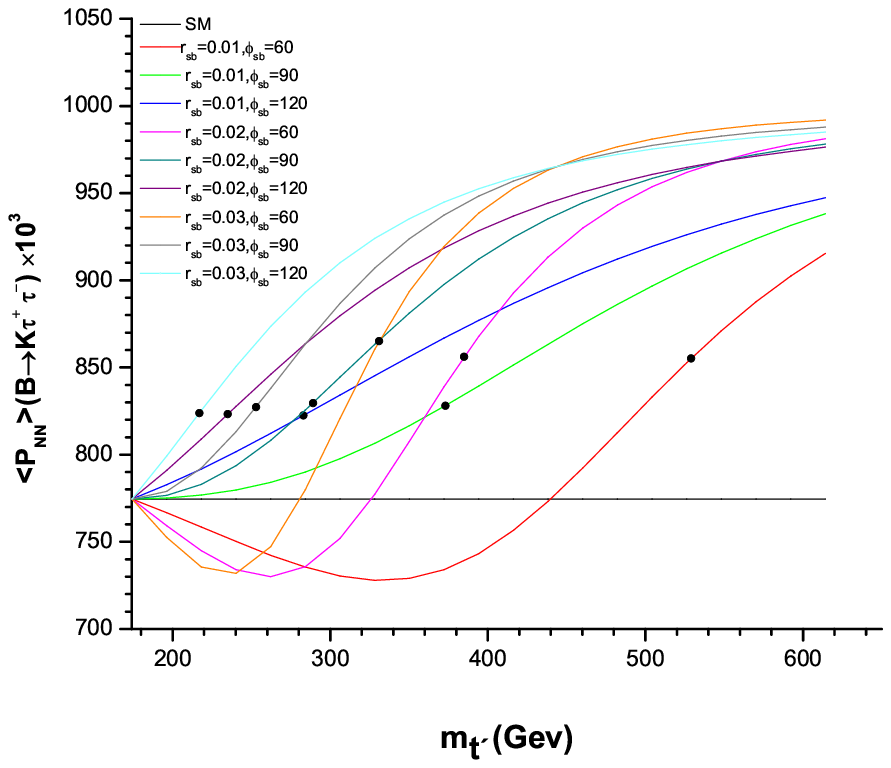}
\vskip 7.8cm \caption{}
\end{figure}

\begin{figure}
\vskip 2.5 cm
    \includegraphics{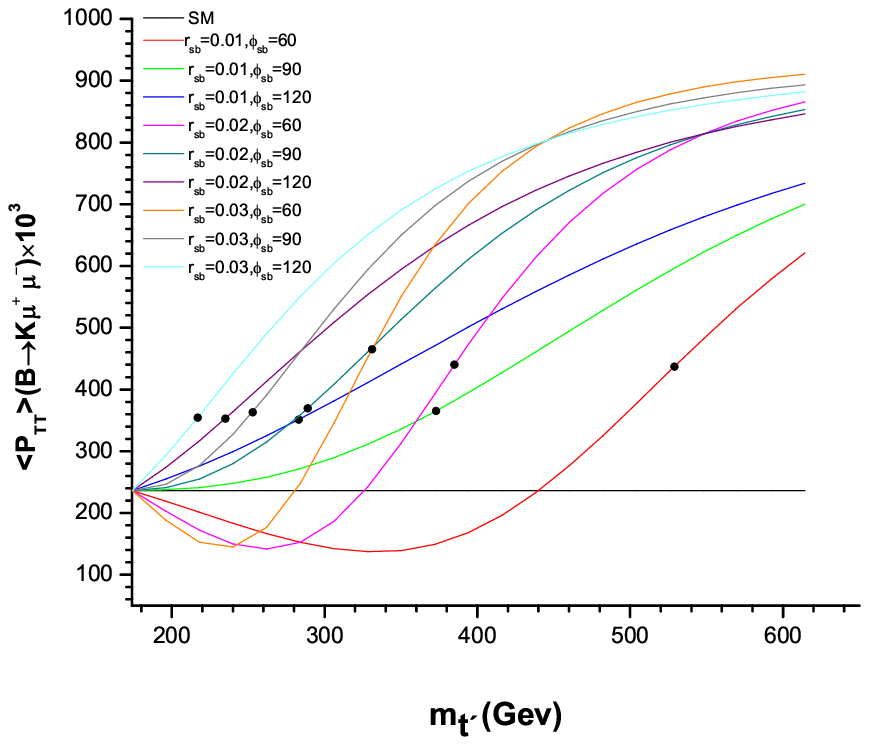}
\vskip 7.8 cm \caption{}
\end{figure}
\include{falahatifig1}

\include{falahatifig2}

\end{document}